\documentclass[preprint,12pt]{elsarticle}

\usepackage{amssymb}
\usepackage{subfigure}
\usepackage{caption}

\journal{}

\begin{document}

\begin{frontmatter}



\title{Effects of human dynamics on epidemic spreading in C\^{o}te d'Ivoire}


\author{Ruiqi Li$^*$\corref{cor1}, Wenxu Wang, Zengru Di$^{**}$\corref{cor2}}
\cortext[cor1]{liruiqi@mail.bnu.edu.cn}
\cortext[cor1]{zdi@bnu.edu.cn}

\address{School of Systems Science, Beijing Normal University, Beijing 100875, PRC}

\begin{abstract}
Understanding and predicting outbreaks of contagious diseases are crucial to the development of society and public health, especially for underdeveloped countries. However, challenging problems are encountered because of complex epidemic spreading dynamics influenced by spatial structure and human dynamics (including both human mobility and human interaction intensity). We propose a systematical model to depict nationwide epidemic spreading in C\^{o}te d'Ivoire, which integrates multiple factors, such as human mobility, human interaction intensity, and demographic features. We provide insights to aid in modeling and predicting the epidemic spreading process by data-driven simulation and theoretical analysis, which is otherwise beyond the scope of local evaluation and geometrical views. We show that the requirement that the average local basic reproductive number to be greater than unity is not necessary for outbreaks of epidemics. The observed spreading phenomenon can be roughly explained as a heterogeneous diffusion-reaction process by redefining mobility distance according to the human mobility volume between nodes, which is beyond the geometrical viewpoint. However, the heterogeneity of human dynamics still poses challenges to precise prediction. 
\end{abstract}

\begin{keyword}
epidemic spreading dynamics \sep human mobility \sep human interaction intensity \sep mobility distance



\end{keyword}

\end{frontmatter}


\section{Introduction}
Understanding and predicting outbreaks of contagious diseases are crucial to the development of society and public health. The degree of risk of outbreaks of some epidemics is quite high in underdeveloped countries, such as bacterial diarrhea (e.g., cholera), SARS, avian flu, H1N1/9, yellow fever and influenza, among others \cite{website:factbook}. Apart from some severe diseases (such as SARS, Ebola), even cholera is a leading cause of death and affects infants and children in low-income countries, in particular \cite{WHOdisease}. Unfortunately, epidemic spreading is becoming increasingly more complex. Hundreds of years ago, epidemic spreading was modeled as a diffusion process with a speed of 300-600 km/year \cite{noble1974geographic}, for which it was quite easy to make precise predictions. However, at present, epidemics spread much faster and can spread around the world in about 6 months \cite{hufnagel2004forecast}, presenting a much more complex phenomenon -- contagious diseases may suddenly appear in a region spatially distant from the original outbreak area, and its path has become difficult to predict.

Evidences increasingly show that epidemic spreading dynamics are influenced by spatial structure and human dynamics (including both human mobility and human interaction intensity) \cite{ni2009impact,watts2005multiscale,VesNP2011phase,gattoPNAS2012generalized,brockmann2009humanReview,brockmann2013hiddenGeo,schlapfer2014scaling,wesolowski2015impact,wesolowski2015quantifying}. With rapid transportation, human mobility has already changed the 2D geometrical space to a higher-dimensional manifold. A place spatially far away can be ``drawn'' much closer by a variety of rapid transit methods (such as airlines, high-speed rails, etc., for inter-city; or light rails, subways, Bus Rapid Transits (BRTs), etc., for intra-city), which means that people living far way geometrically may not be really distant. In this perspective, the world we perceive is not how the real world ought to look. This change will strongly affect epidemic spreading dynamics. Additionally, there is a super-linear relationship between the total human interactions and city size \cite{schlapfer2014scaling}, which indicates that epidemics have more opportunity to spread in larger cities and with different infection durations. The temporal aspect of human dynamics, especially the bursty nature of human contacts, is also very important. Great efforts have been devoted to determine the impact of the temporal aspect of human dynamics on epidemic spreading \cite{vazquez2007impact,iribarren2009impact,yang2011impact,karsai2011small,min2011spreading}. These factors all pose great challenges to epidemic modeling and prediction.

Therefore, it is of great importance to develop a realistic dynamic model that incorporates the effects of spatial structure and human dynamics (both human mobility and human interaction intensity) to understand and predict the spreading patterns and provide insights into the course of past and ongoing epidemics (especially in low-income countries), which can be helpful in assisting in emergency management and allocating health-care resources via an assessment of intervention strategies \cite{rinaldo2012reassessment,bertuzzo2011prediction,mari2011modelling,tuite2011cholera,chao2011vaccination,mukandavire2011estimating}.

In this paper, we mainly focus on the effects of human mobility and human interaction intensity on contagious diseases spreading in C\^{o}te d'Ivoire at sub-prefecture spatial resolution based on call detailed record (CDR) data provided by the Orange Group and organizers of the D4D Challenge \cite{D4D2012Cote,website:OrangeInc}. More specifically, we concentrate on modeling and predicting the epidemic spreading process once cases appear in one sub-prefecture and providing a systematic analysis of the conditions under which an epidemic can start and along what kind of path it spreads. Interestingly, the spreading process is not a classical diffusion process but is a kind of l\'{e}vy flight \cite{brockmann2009humanReview}, which consists of local diffusion and long-range jumps. Furthermore, we determine the underlying mechanism behind such kinds of l\'{e}vy flight. By defining a proper distance, beyond the geometrical distance, according to the human mobility matrix at sub-prefectures resolution, we can recover the diffusion process from the l\'{e}vy flight. Apart from the ``effective distance'' proposed in \cite{brockmann2013hiddenGeo}, the inverse of the human mobility volume is also a proper distance for predicting the spreading process. The effects of heterogeneous human interaction intensity, which can lead to some repetitive infected nodes, should also be considered when defining a proper distance.

The remainder of this paper will be organized as follows. Section II will introduce our model and the associated improvements compared with former models and the results. Section III presents the conclusion and discussion. The Appendix will mainly describe the details of the CDR data set \cite{website:OrangeInc}, how we extract human mobility and interaction intensity from it, and other related empirical results which support our model.

\section{Model and Results}
\begin{figure}   \centering
\includegraphics[width=4.5in]{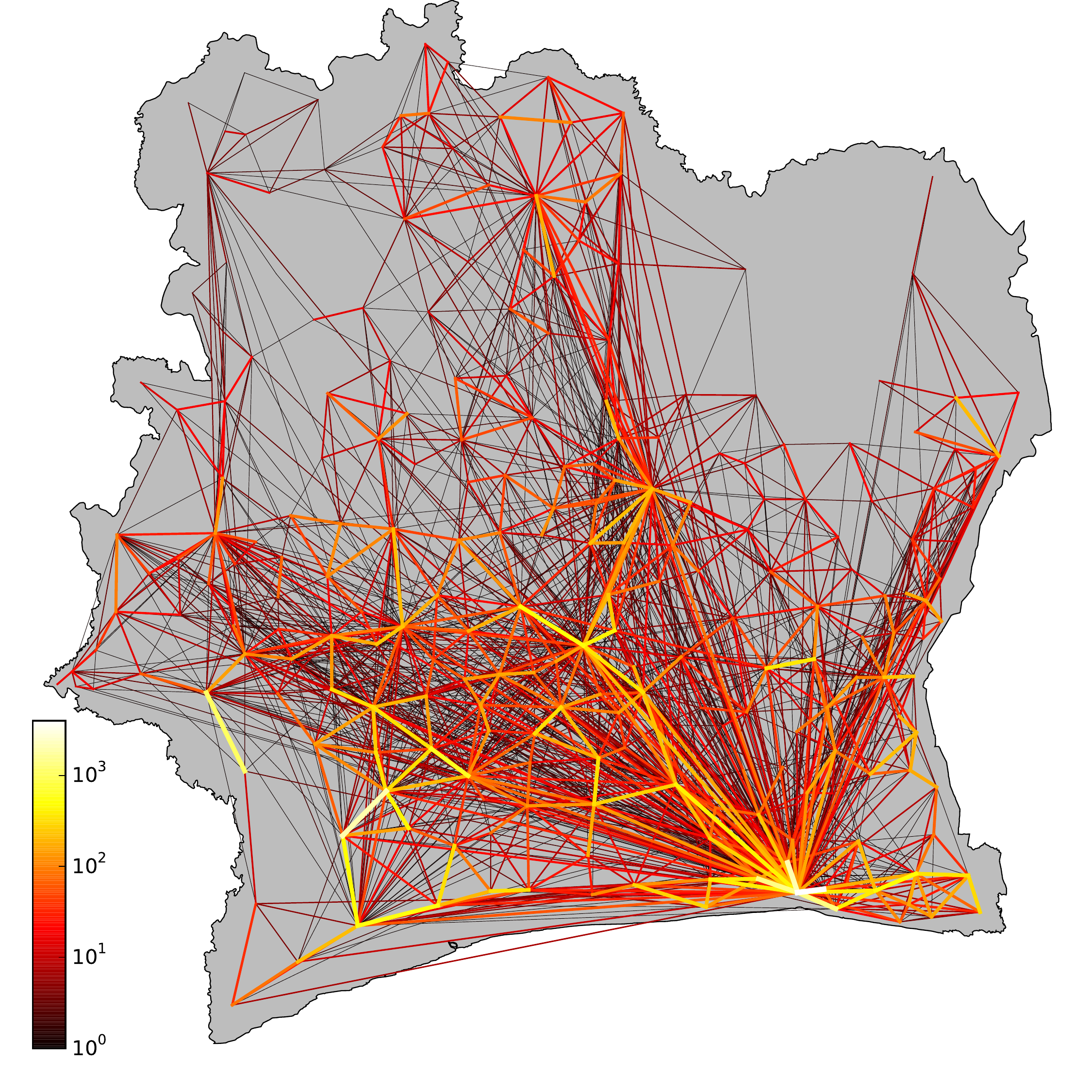}
\caption{The average daily human mobility volume between sub-prefectures. A lighter and thicker line indicates a higher mobility volume. The human mobility matrix illustrated in this figure is the average result of the total mobility volume during 150 days.}\label{fig.flow}
\end{figure}

The basis of our analysis is a spatially explicit nonlinear differential model that incorporates mainly human mobility networks and human interaction intensity. The epidemic spreading dynamic is a classical susceptible-infected-recovered (SIR) contagion process in metapopulation networks \cite{ni2009impact,watts2005multiscale,VesNP2011phase,wesolowski2015impact,wesolowski2015quantifying,keeling2008modeling}. There are 255 nodes in our model, corresponding to 255 sub-prefectures in C\^{o}te d'Ivoire, and 21,952,093 individuals (the total estimated population by CIA \cite{website:factbook} in 2013) are allocated within these nodes, according to the population distribution (see the Appendix for details). In the $i$th sub-prefecture, with a population of $N_i~(i=1,2,...,255)$, the state variables at time $t$ are the local abundances of susceptible individuals, $S_i(t)$, infected individuals, $I_i(t)$, and recovered people, $R_i(t)$. The population distribution (i.e., $N_i$) is estimated based on the census in 1989 (see the Appendix for details). Connections between sub-prefectures are characterized by the human mobility volume, which can be denoted as a matrix $A=[A_{ij}]$, where $A_{ij}$ is the number of people moving from sub-prefecture $i$ to $j$, extracted from the CDR data (see Fig. \ref{fig.flow} and the Appendix for details). The CDR data provide us the anonymous user-id, timestamp, subpref-id (and its position) when the user makes a call. Once the user makes a call in a new place, we can identify a movement (since the resolution is at sub-prefecture, this movement extraction is little affected by the signal noise (i.e., the people doesn't move but due to tower traffic balancing, the signal might be assigned to another tower)\cite{jiang2013review} , see Appendix for more details of movement extraction). Using real human mobility data is an advantage compared with previous models, where the mobility volume between cities was estimated by gravity or a gravity-like model \cite{VesNP2011phase,gattoPNAS2012generalized,rinaldo2012reassessment,bertuzzo2011prediction,mari2011modelling,tuite2011cholera,chao2011vaccination,mari2012role}.

\begin{figure}   \centering
\includegraphics[width=4.5in]{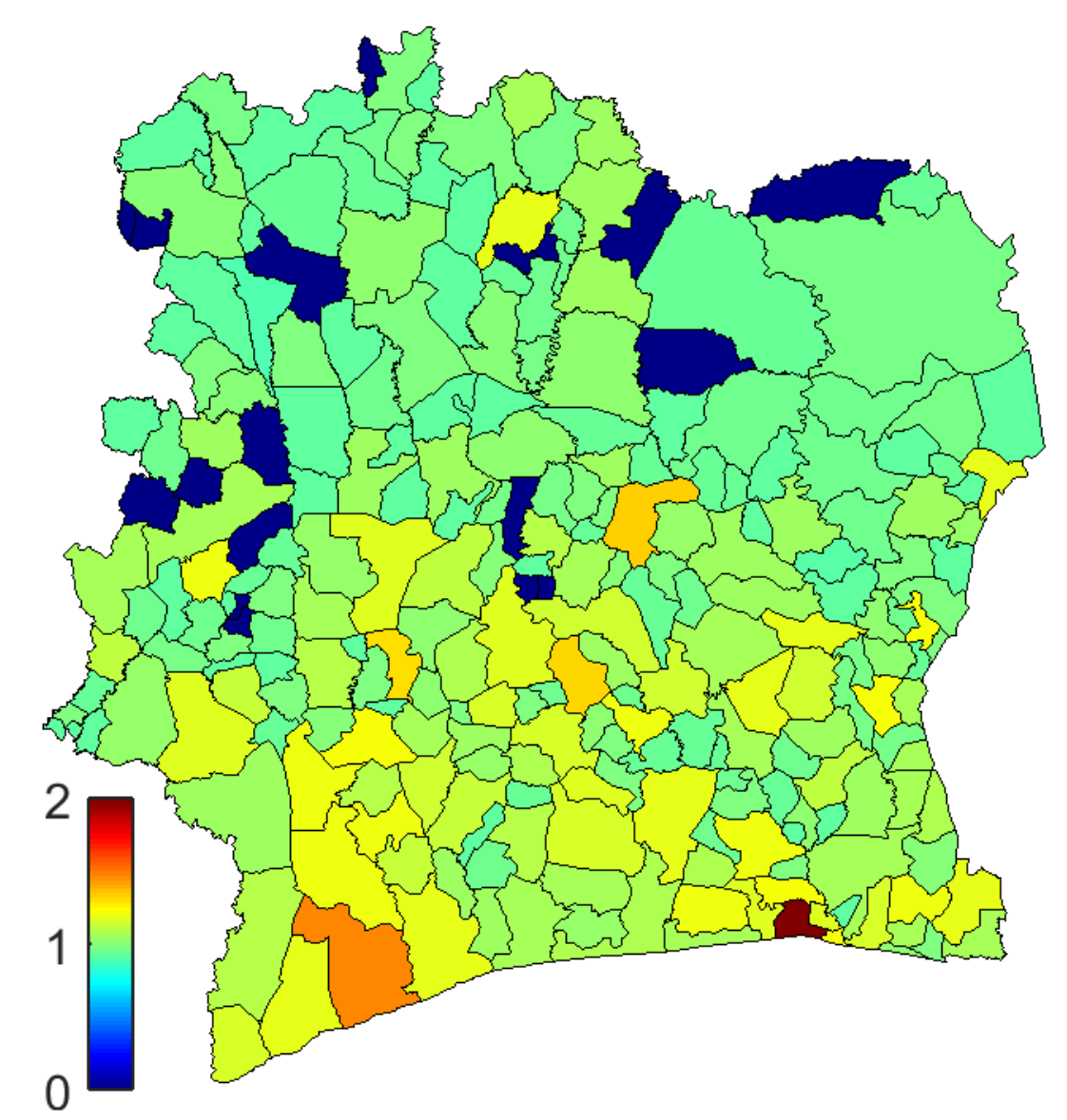}
\caption{The spatial distribution of $R_0$ at sub-prefecture resolution when $\langle R_0\rangle=1$. There are 18 sub-prefectures with no population or communication volume; we set their BRNs to 0.}\label{fig.BRN}
\end{figure}

\begin{figure}  
\includegraphics[width=5.5in]{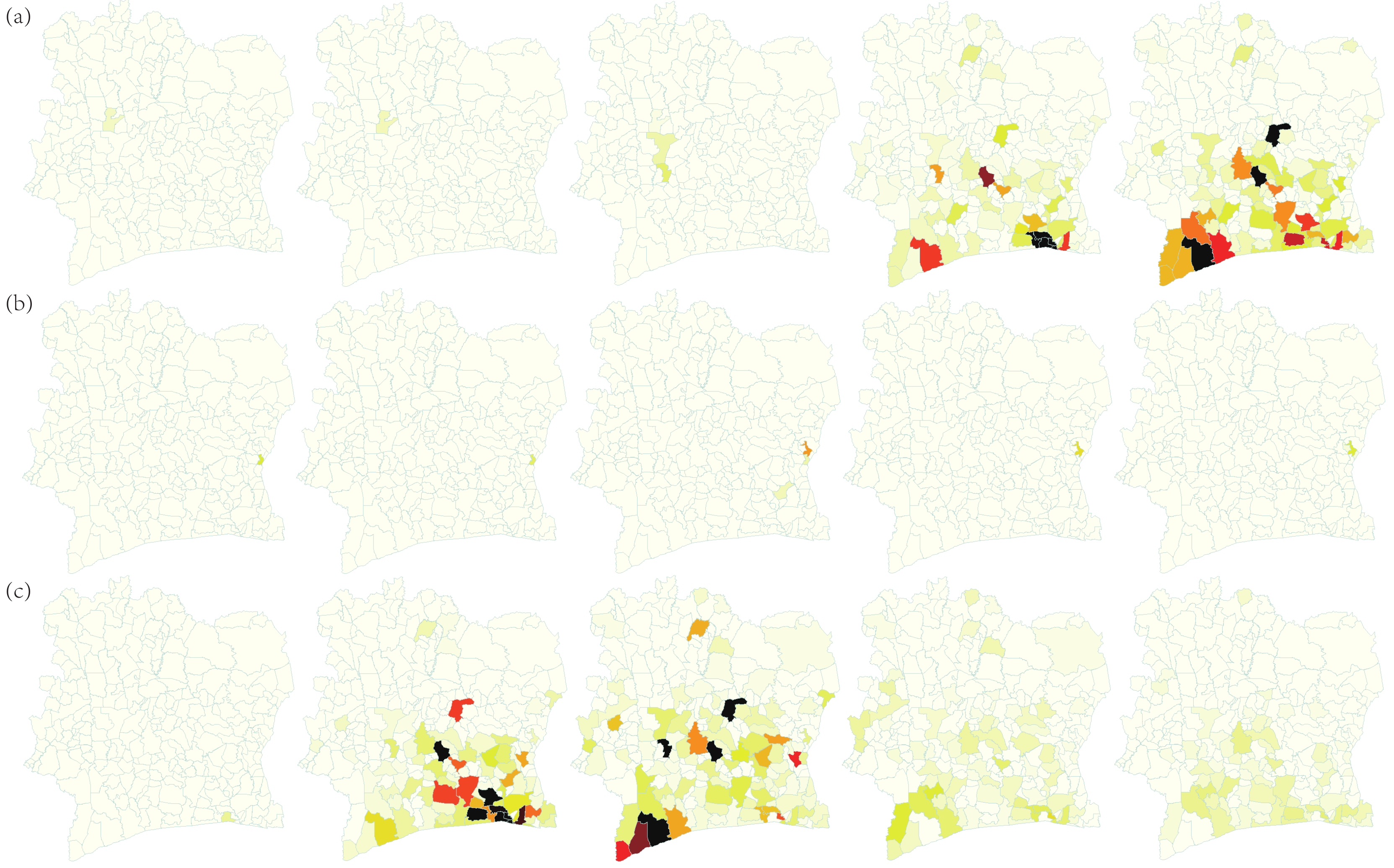}
\caption{An illustration of the epidemic spreading process occurring in a certain sub-prefecture with 10 infected people. We choose five snapshots of day 1, 40, 80, 120 and 150, respectively. The source outbreak regions are (a) Seguela, (b) Akoboissue, and (c) Abidjan.}\label{fig.spread}
\end{figure}

In order to quantitatively depict the effects of interaction intensity upon the infectivity of an infectious disease, we introduce the local basic reproductive number (BRN) $R_0$, which is the number of infected cases that one case generates on average over the course of his infectious period. When $R_0<1$, the infection will die out in the long run in that region; if $R_0>1$, the infection will be able to spread among the population. The local BRN of sub-prefecture $i$ is
\begin{equation}
  R_{0i}=\frac{S_i\eta_i\beta_i}{\mu_i},
\end{equation}
where in sub-prefecture $i$, $S_i$ is the susceptible population and $\eta_i$ and $\beta_i$ are the probabilities of a susceptible individual having contact with and getting infected by an infected individual, respectively. $\mu_i$ is the sum of death baseline (i.e., $\mu_{1i}$) and recovery probability (i.e., $\mu_{2i}$). For $\beta_i$ and $\mu_i$, due to various elements (such as climate, sanitary, hydrological conditions, etc.), there might be some slight deviations among the sub-prefectures. These parameters can be measured or estimated. We assume
\begin{equation}
  \eta_i=1-(1-\frac{I_i}{N_i})^{x_i+h_i}\approx \frac{(x_i+h_i)I_i}{N_i},
\end{equation}
which is positively proportional to the local household size $h_i$ and the local social interaction intensity $x_i$, because under the usual conditions, a user will have contact with his or her family members and also have social activities outside,. $h_i$ is determined from the census \cite{website:factbook}, and $x_i$ is estimated according to a super-linear relationship with population size $N_i$ ($x_i\approx N_i^{1.12}$ according to \cite{schlapfer2014scaling}). For example, the level of local interaction intensity in Abidjan (economical and previous capital) will be much higher than in remote regions, which on average means that the people in Abidjan will have contact with more people apart from their family members compared to people in remote sub-prefectures.
After introducing an infected individual into a completely susceptible population, the number of new infections per unit time (e.g., one day) is $S_i\eta_i\beta_i$. This expression has to be multiplied by the average length of the infectious period $1/\mu_i$, which leads to
\begin{equation}
  R_{0i}=\frac{S_i(x_i+h_i)I_i\beta_i}{N_i\mu_i}.
\end{equation}
In the onset phase, $S_i\approx N_i,~I_i\approx 1$, which gives us $R_{0i}\approx (x_i+h_i)\beta_i/\mu_i$. We assume that average BRN $\langle R_0\rangle=\sum_i R_{0i}/N$ gives us an overview of the whole country, where $N$ is the number of sub-prefectures in C\^{o}te d'Ivoire. When $\langle R_0\rangle=1$, the distribution of $R_{0i}$ is shown in Fig. \ref{fig.BRN}, which ranges from about 0.87 to 1.91. According to classical epidemiological theory \cite{bailey1975the,anderson1991infectious}, when $\langle R_0\rangle>1$, the epidemic will prevail rather than die out. Although the BRN is a good qualitative tool for estimating the condition of a sub-prefecture, when incorporated with human mobility, it might not be that correct when applied at the nation scale \cite{gattoPNAS2012generalized}. For those sub-prefectures with $R_{0i}$ greater than one, an endemic will certainly occur. Whether it will have a chance to trigger a nationwide epidemic is then mainly determined by human mobility and the susceptibility of connected regions. For example, if there is a disease outbreak in Seguela (sub-prefecture200 in the data), although its BRN is less than 1, it still triggers a nationwide epidemic mainly due to its connection to some well-connected regions with high risk (See Fig. \ref{fig.spread}(a) and Supplementary movie 1). On contrast, if it first appears in Akoboissue (subprefecture1 in the data), which has a  BRN greater than 1, it only spreads locally due to the sparse connectivity to other regions which may also be with lower susceptibility (see Fig. \ref{fig.spread}(b) and Supplementary movie 2). The requirement that the average BRN $\langle R_0\rangle$ to be greater than unity is not necessary for outbreaks of epidemics.

\begin{figure}   \centering
\includegraphics[width=4.5in]{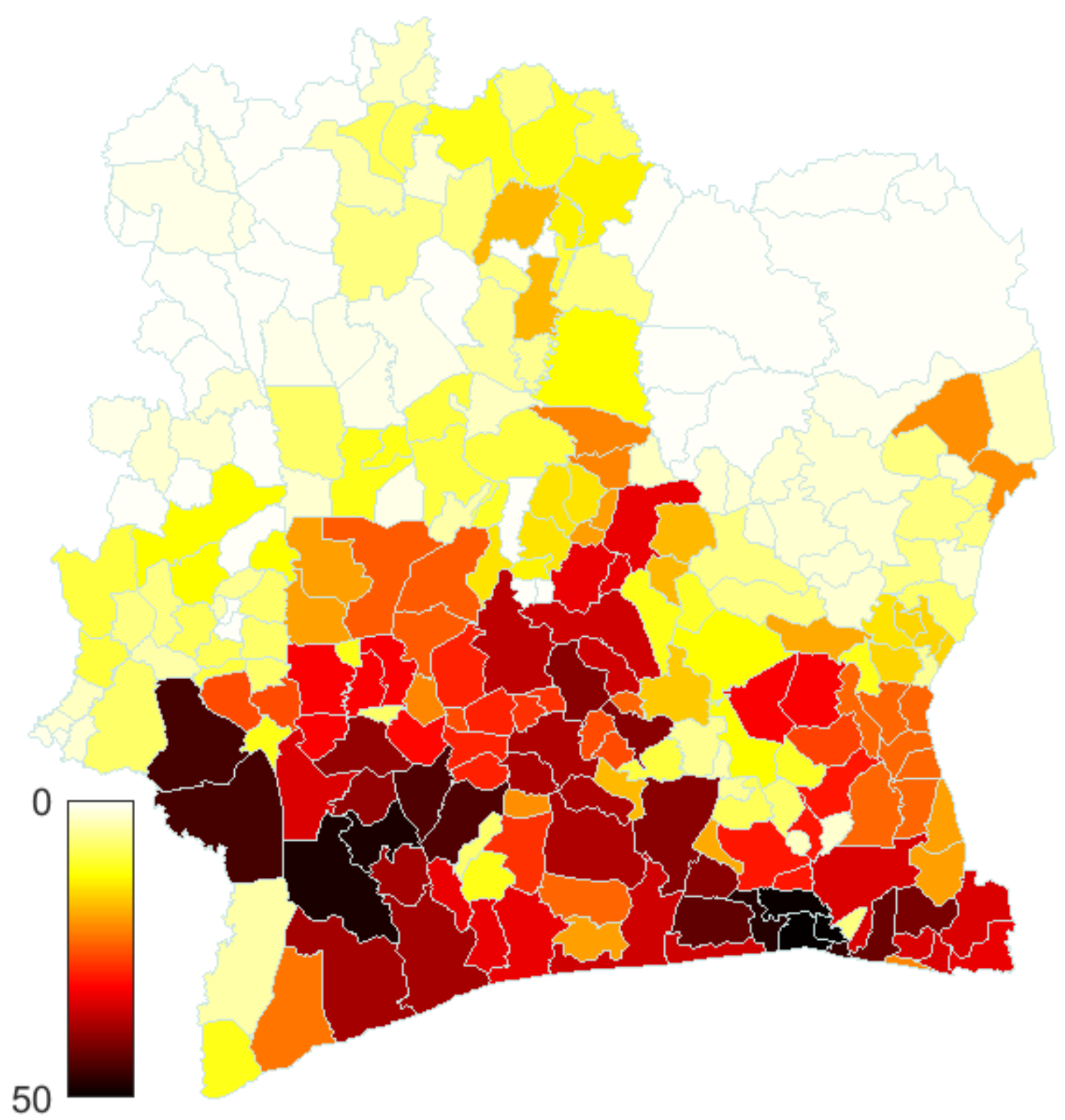}
\caption{Results of the weighted k-shell decomposition at the sub-prefecture level. We neglect links with a weight less than 10 (we divide all the link weights by 10, i.e., we set 10 as the volume basis).}\label{fig.kshell}
\end{figure}

In order to depict the impact of human mobility, we use a weighted k-shell method \cite{weightedkshell,Makse2010kshell,PRL2006kcore} to detect the potential regions based on the mobility matrix. We consider the unitary weight for a link to be 10, i.e., we get the weight of the links by dividing by 10 and just considering the integer part. We start by iteratively removing the nodes with the minimum weighted degree. After the first round, if there are some nodes with a weight not greater than the previous minimum weighted degree, we also remove them. We repeat this process until we end up with a complete graph with same weight, i.e., the central core. Surprisingly, the core does not just include the Abidjan regions but also consists of two regions quite far apart (see Fig. \ref{fig.kshell}). Interestingly, the nodes with the same k-shell value seem to be distributed very randomly, which indicates that tightly connected regions are not necessarily concentrated geometrically. There is some evidence that the activation of the epidemic is mainly due to the hub node (or core) staying in the infected state \cite{castellano2012competing}; therefore, if we can intervene at these nodes or the connections to these regions, the epidemic may not become that serious. In addition, some recent works have discovered that in some real-world networks, there are core-like groups that have high k-shell values but are just locally connected clusters rather than the real core of the network, which can be detected by link entropy \cite{liu2015core}. After removing some redundant links (the redundancy is defined by the spreading influence), the k-shell value will provide a more precise evaluation of the core centrality \cite{liu2015improving}.

After considering the effects of interaction intensity and human mobility, we integrate these two important factors into our model, which can be expressed as
$$\left\{
\begin{array}{ccl}
  \frac{dS_i}{dt}&=&-\mu_1S_i-(S_i-S_i^{out})\eta_i\beta-\sum_{j=1,j\neq i}^{255}S_{ij}^{out}\eta_j\beta \\
  \frac{d I_i}{d t}&=&-(\mu_1+\mu_2)I_i+(S_i-S_i^{out})\eta_i\beta+\sum_{j=1,j\neq i}^{255}S_{ij}^{out}\eta_j\beta \\
  \frac{d R_i}{d t}&=&\mu_1S_i+(\mu_1+\mu_2)I_i
\end{array},
\right. $$
where the susceptible out-flow of node $i$ is $S_i^{out}=A_{ij}S_i/N_i$ and susceptible out-flow from node $i$ to node $j$ is $S_{ij}^{out}=(A_{ij} S_i)/N_i,~\eta_i=1-(1-I_{i_t}/N_i )^{x_i+h}\approx (x_i+h)I_{i_t}/N_i$. If we assume that the mobility pattern of the infected agent (e.g., getting a cold or cholera, etc.) is not affected by the disease, then we obtain that the infected population of $i$ on a certain day is $I_{i _t}=I_i-\sum_{j\neq i}A_{ji}(\rho_j-\rho_i)$, where $\rho_i=I_i/N_i$. Therefore, within node $i$, the probability of contacting with infected agents $\eta_i$ is affected by the infectious population there and the difference between infectious densities with other nodes. In order to highlight the effects of human dynamics upon the disease spreading, we neglect slight differences in some elements (we degenerate some diagonal matrices to constants, i.e., $\beta_i=\beta,~h_i=h,~\mu_{1i}=\mu_1,~\mu_{2i}=\mu_2$).

\subsection{Data-driven Simulation}
\begin{figure}   \centering
\includegraphics[width=4.5in]{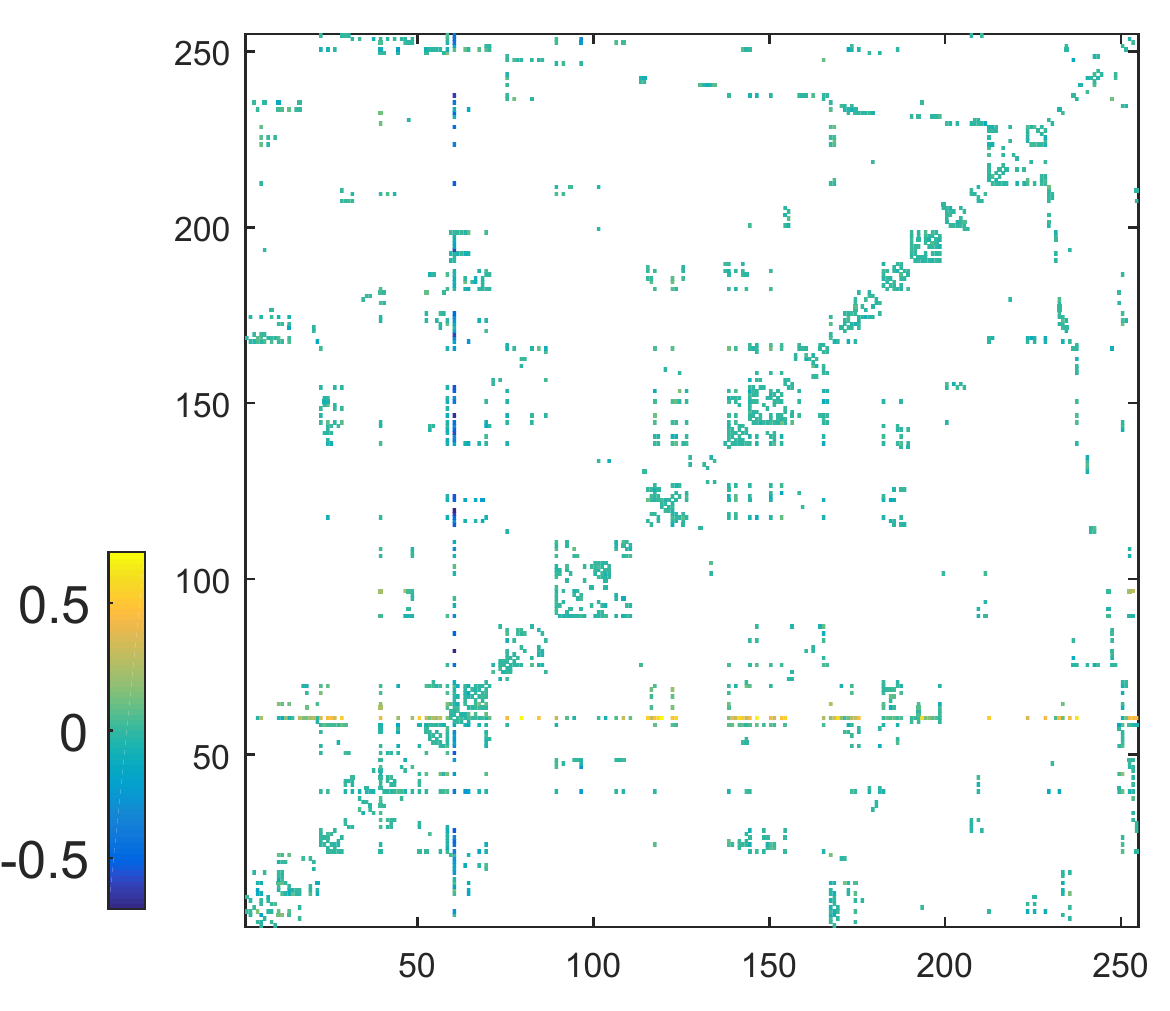}
\caption{Conservation matrix $B[B_{ij}]=[2(A_{ji}-A_{ij})/(A_{ij}+A_{ji})]$ of the mobility flow. Apart from Abidjan (serial number 60), the mobility flow between sub-prefectures is conservative. We neglect the node pair with an average daily volume less than 2.}\label{fig.conserve}
\end{figure}

We extract the mobility network at sub-prefecture level from the CDR data \cite{D4D2012Cote,website:OrangeInc}. In the simulation, each sub-prefecture is associated with its estimated population (see Appendix). The population may fluctuate due to human mobility; however, by defining a balancing matrix $B=[B_{ij}]=[2(A_{ji}-A_{ij})/(A_{ij}+A_{ji})]$, where $A_{ij}$ is the human mobility volume from node $i$ to node $j$, if there are more people from $i$ going to $j$ than $j$ to $i$, then $B_{ij}$ is negative, which mean $i$ is ``losing'' people, and vice versa. We found that the in- and out-flow are almost conserved apart from the economical capital Abidjan, which may due to the greater number of opportunities and jobs there (see Fig. \ref{fig.conserve}). We assume that all the people who go to other places will return to their homes; thus, the population of each sub-prefecture remains unchanged. The parameters are all determined in the same way as previously stated in the model, and non-human-dynamic related parameters are degenerated to be constants, as well.

We select one sub-prefecture as the infectious source and initiate it to have 10 infected people. The mobility volume $A_{ij}$ is updated daily for the total 150 time steps. Based on real mobility and human interaction data, we have simulated the spreading of an SIR contagion process and studied the number of infected sub-prefectures and the number of infected populations and the first arrival time of each sub-prefecture. We choose three typical spreading process as demos in Fig. \ref{fig.spread} and Supplementary Movies 1--3 --- Seguela (sub-prefecture 200), Akoboissue (sub-prefecture 1), Abidjan (sub-prefecture 60).

The corresponding analysis shows that the impacts caused by human mobility on epidemic dynamics are really strong -- even for the sub-prefecture with a BRN less than 1 (where the endemic will not prevail locally), the human mobility flow can bring the disease to some other places, which may trigger nationwide epidemic outbreaks (e.g., Seguela, see Fig. \ref{fig.spread}(a) and Supplementary Movie 1). While there is no reason to worry about such kinds of cascading effects in certain sub-prefectures due to the low volume of human mobility flow (e.g., Akoboissue, see Fig. \ref{fig.spread}(b) and Supplementary Movie 2), it can only affect a few sub-prefectures connected to it, while other places are safe. In general, due to the effects of human mobility, the spreading process is no longer a classical diffusion process around the source geometrically but is rather a kind of l\'{e}vy  flight: if we set Abidjan as the source of the epidemic, we can observe that it does not affect all the surrounding regions first, but infect some sub-prefectures located in the middle and north in a short period, exhibiting strong spatial heterogeneity (see Fig. \ref{fig.spread}(c) and Supplementary Movie 3). The dynamical evolution pattern differs greatly from the classical conclusions. We show that the requirement that the average BRN $\langle R_0\rangle$ to be larger than unity is not necessary for epidemic outbreaks when local settlements are connected by mobility networks of primary and secondary infection mechanisms. Combining the human interactions and weighted k-shell results, we would be able to identify the critical paths which connect the high-risk regions (i.e., $R_{0i}>1$) and the core of the network. Then, by proper intervening strategies, we might be able to prevent some nationwide epidemic outbreaks. For example, in Fig. \ref{fig.spread}(a) and Supplementary Movie 1, if we restrict or make more surveillance on the mobility between sub-prefecture 60 and the sub-prefecture right below it, we can avoid an epidemic outbreak. 

\begin{figure}  \centering
\includegraphics[width=5in]{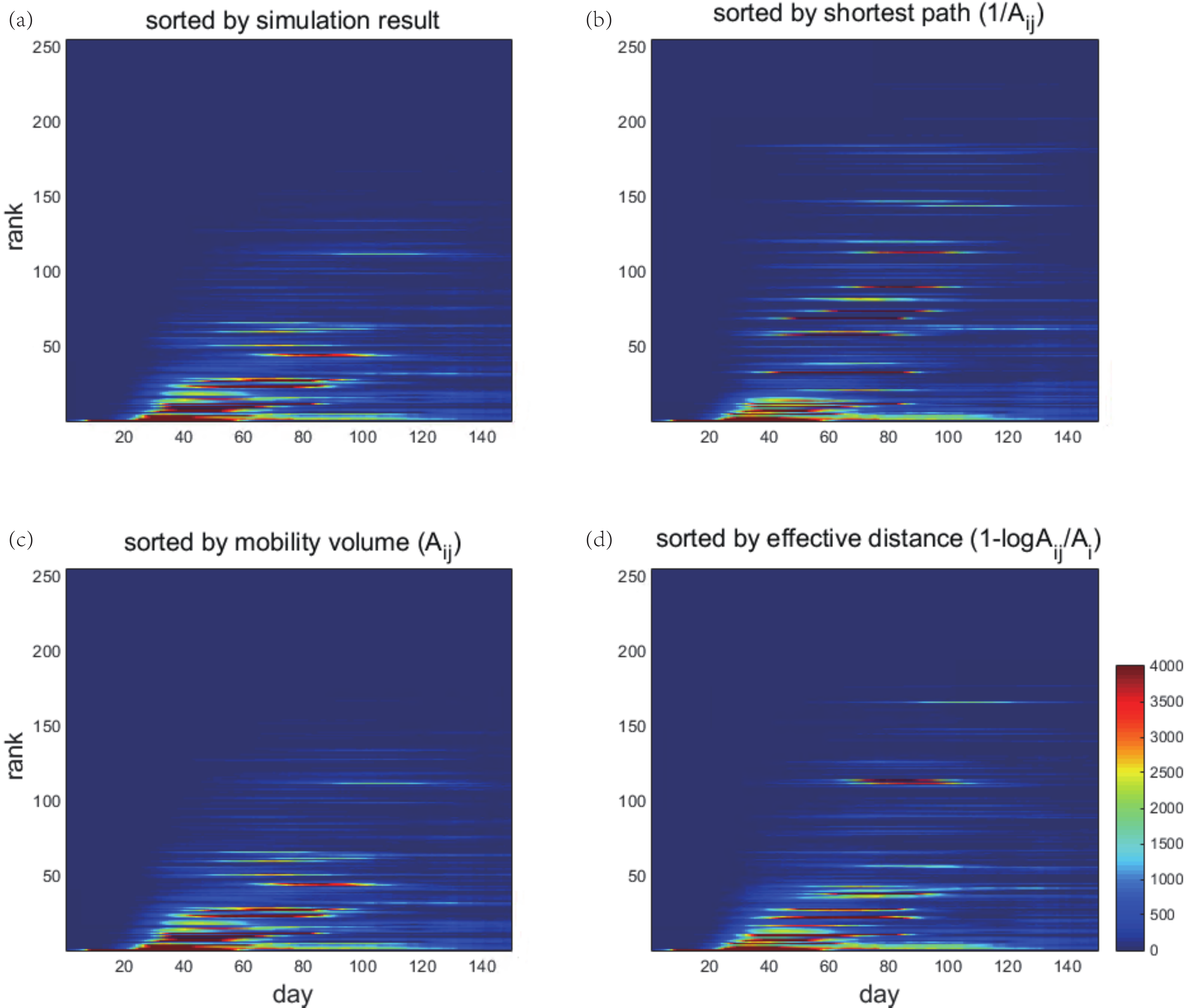}
\caption{The temporal view of the epidemic spreading process nationwide. The first outbreak source node is Abidjan. The $X$ axis shows the time, and the $Y$ axis is the index of sub-prefectures ranked by (a) the first arrival time in the simulation; (b) the shortest paths to the source node by assuming $d_{ij}=1/A_{ij}$; (c) the mobility volume $A_{60j}$ from the source node to the others; (d) ranked by the effective distance $d_{ij}=1-\log(A_{ij}/A_i$) \cite{brockmann2013hiddenGeo} from the source node to the others.} \label{fig.pcolor}
\end{figure}

\begin{figure}   \centering
\includegraphics[width=5in]{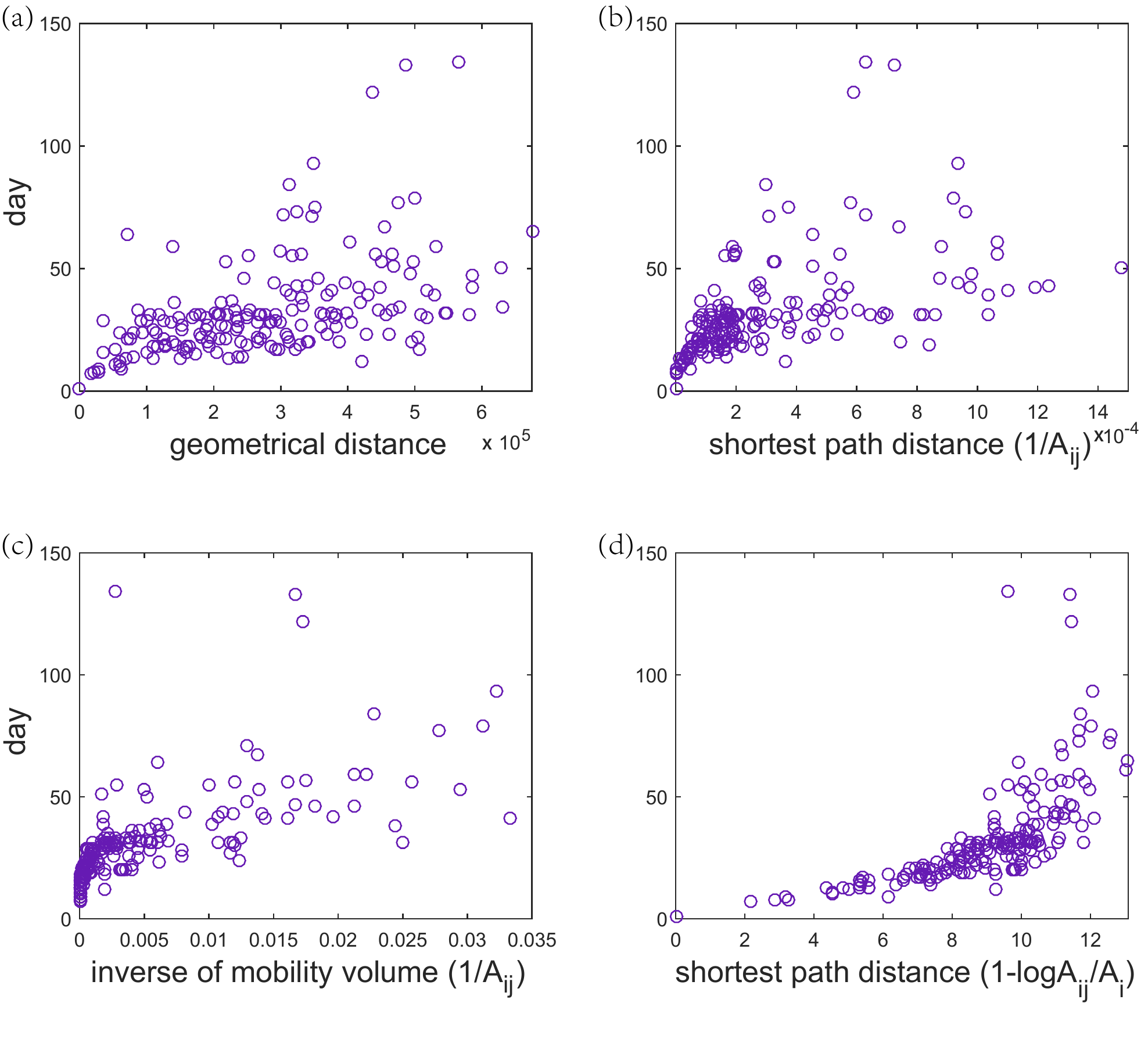}
\caption{The first arrival times for each sub-prefecture and (a) the geometrical distance (in meters) to the source node; (b) the shortest path length defined by $d_{ij}$ as $1/A_{ij}$; (c) the mobility volume $A_{ij}$ from the source node to the others; (d) the shortest path length defined by $d_{ij}$ as $1-\log(A_{ij}/A_i$). For better visualization, in (b) and (c), about 6 nodes with very large magnitudes are not shown (see Fig. \ref{fig.Ascatter} in the Appendix.)}\label{fig.scatter}
\end{figure}

There is some evidence showing that the observed l\'{e}vy flight phenomena can be mapped back to a classical diffusion process when the geometrical distance is not considered but an ``effective distance'' is defined according to human mobility \cite{brockmann2013hiddenGeo}. For example, although some sub-prefectures are far away from Abidjan geometrically, the mobility volume between them can be very high, and from the effective distance viewpoint, they are very close and always have strong influences on each other.
According to this kind of effective distance, we are able to predict the first arrival time of the epidemic for each node. We compare the results of the simulation (see Fig. \ref{fig.pcolor}(a) and Fig. \ref{fig.scatter}(a)) with three types of distances $d_{ij}$ defined by human mobility. The first one is the inverse of the mobility volume $d_{ij}=1/A_{ij}$, and we then calculate the shortest path from the source node to others and rank the nodes in ascending order (see Fig. \ref{fig.pcolor}(b) and Fig. \ref{fig.scatter}(b)). The second one is just the mobility volume $d_{ij}=A_{ij}$, and we rank the nodes in descending order (see Fig. \ref{fig.pcolor}(c) and Fig. \ref{fig.scatter}(c)). The third one is the effective distance ($d_{ij}=1-\log(A_{ij}/A_i)$) proposed in \cite{brockmann2013hiddenGeo} sorted in ascending order (see Fig. \ref{fig.pcolor}(d) and Fig. \ref{fig.scatter}(d)).

In Fig. \ref{fig.scatter}, for the quantitative description, the geometrical distance to the source node is not appropriate for predicting the spreading (see Fig. \ref{fig.scatter} (a)). The shortest path of the inverse of the mobility volume is better for use when the distance is not too large (see Fig. \ref{fig.scatter} (b)). The prediction based on the mobility volume $A_{ij}$ is too concentrated in the first region and dispersed in the latter region (see Fig. \ref{fig.scatter} (c)). The effective distance does not provide a good prediction for those nodes that are ``far away'' from the source node as well.
In a densely connected core (see Fig. \ref{fig.kshell}), the first arrival times of many nodes are quite close in value, which means that in the later spreading process, there will be more paths that can spread the disease. The most strongly (densely) connected network is the complete graph. From an inverse spreading perspective \cite{shen2014reconstructing,han2015robust}, if we want to determine the source of the infection, we will need to determine the infection time sequence of almost all the nodes; as for the most sparse but connected situation, i.e., a chain, we can infer the position of the source node with information on any two nodes in the network. Additionally, from Fig. \ref{fig.pcolor} and Fig. \ref{fig.Apcolor} in the Appendix, we can also clearly observe that the duration and intensity of the disease at each place are quite heterogeneous. Evidence has shown that the activation of the epidemic is mainly due to the hub node or the hub core remaining in the infected state \cite{castellano2012competing}. This phenomenon also violates the branching process that poses challenges in terms of predicting the spreading path. The outbreak intensity within a place may also introduce some complexity.

\section{Conclusion and Discussion}
In this paper, we propose a more realistic model for epidemic spreading based on human mobility and human interaction intensity, which can provide us a qualitative and quantitative depiction of the dynamical process. Considering the human mobility matrix, we show that the requirement that the average BRN $\langle R_0 \rangle$ is to be larger than unity is not necessary for nationwide epidemic outbreaks, which is strongly affected by human dynamics. By abandoning the well-mixed assumption and considering the heterogeneity of human interaction intensity, we can observe more realistic spatiotemporal patterns of epidemic spreading and that the infection duration and intensity of epidemic in each place are quite different from each other. These phenomena cannot be captured by previous models and pose challenges in terms of precise prediction. Under the framework of our model, combining the human interactions and weighted k-shell results, we can identify critical paths for preventing nationwide epidemic outbreaks in the future. In addition, by defining a proper distance according to human mobility, we can roughly map the observed l\'{e}vy-flight-like spreading process back to a classical diffusion process and then qualitatively predict the spreading path of the epidemic.
However, two main factors pose challenges in terms of predicting the spreading path. One is the densely connected component in the network: the effects of the interactions between nodes is very complex, which will certainly violate the branching process of epidemic spreading.
Another factor is the heterogeneity in human interaction intensity, leading to different infection duration and intensity for each place, which introduces some repeatedly infected nodes. Integrating the effects of different duration, intensity and human mobility for more precise predictions requires further study.

Our model can be applied in more complicated situations by slightly modifying the connectivity matrix $A$ in our model to incorporate other effects, such as hydrological networks or other similar related factors. For example, we can treat the contaminated water as the directed flow of an appropriate number of infected individuals from one place to another. However, due to limited data accessibility, we still neglect local sanitation, hydrological situation, distribution of schools and workplaces and public places, different means of transportation that can be treated as a multiplex or multi-relational network \cite{rui2013epidemic}, and the intervention strategy (such as immunization, vaccination, school close, case isolation, etc.). Integrating these important factors into a systematical model that can be applied at different scales requires future studies.


\section{Acknowledgements}
We acknowledge the organizers of the D4D Challenge for permitting us to use the C\^{o}te d'Ivoire CDR dataset. This work is supported by NSFC under grant Nos. 61374175 and 11105011.

\appendix
\section{Empirical Study of the Population Distribution}
To the best of our knowledge, in C\^{o}te d'Ivoire, before 2013, no census was conducted at the sub-prefecture resolution since 1989 (a very recent census was conducted in 05/15/2015 \cite{website:CotePop} after we finished our research.). We therefore estimated the population $N_i$ by multiplying a proper scalar to obtain the total estimated population in 2013 based on the census data in 1998 \cite{website:CotePop}. Due to some social and historical problems, 22 sub-prefectures did not exist or were not inhabited in 1998. For these regions, we assigned a population with similar cell phone activity situations. After obtaining the recent census data, we presented the top 10 largest cities for a comparison, and their populations accounted for a large proportion of the population of C\^{o}te d'Ivoire. We find the difference ratios between our estimation after scaling and recent Census are acceptable (see Table \ref{table.pop}).

\begin{table}
\caption{Population comparison of the 10 biggest cities in C\^{o}te d'Ivoire}
\label{table.pop}
\begin{tabular}{lcccc}
\hline
city&	Census(1989)&	After Scaling&	 Census(2015)	& Diff. ratio \\
\hline
Abidjan&	2,877,94&4,246,001 &	4,395,243 &	0.03\\
Bouak¨¦&	461,618 &	681,051 &	542,082 &	0.26\\
Daloa&	173,107 &	255,395 &	266,324 &	0.04\\
Korhogo&	142,039 &	209,558 &	245,239 &	0.15\\
Yamoussoukro&	155,803 &	229,865 &	207,412 &	0.11\\
San-Pedro&	131,800 &	194,452 &	174,287 &	0.12\\
Gagnoa&	107,244 &	158,223 &	167,900 &	0.06\\
Man&	116,657 &	172,111 &	148,171 &	0.16\\
Anyama&	79,548 &	117,362 &	115,260 &	0.02\\
Divo&	86,569 &	127,720 &	105,859 &	0.21\\
\hline
\end{tabular}
\end{table}

\section{Empirical Study of the Human Mobility Pattern Based on CDR}
\begin{figure}   \centering
\includegraphics[width=3.5in]{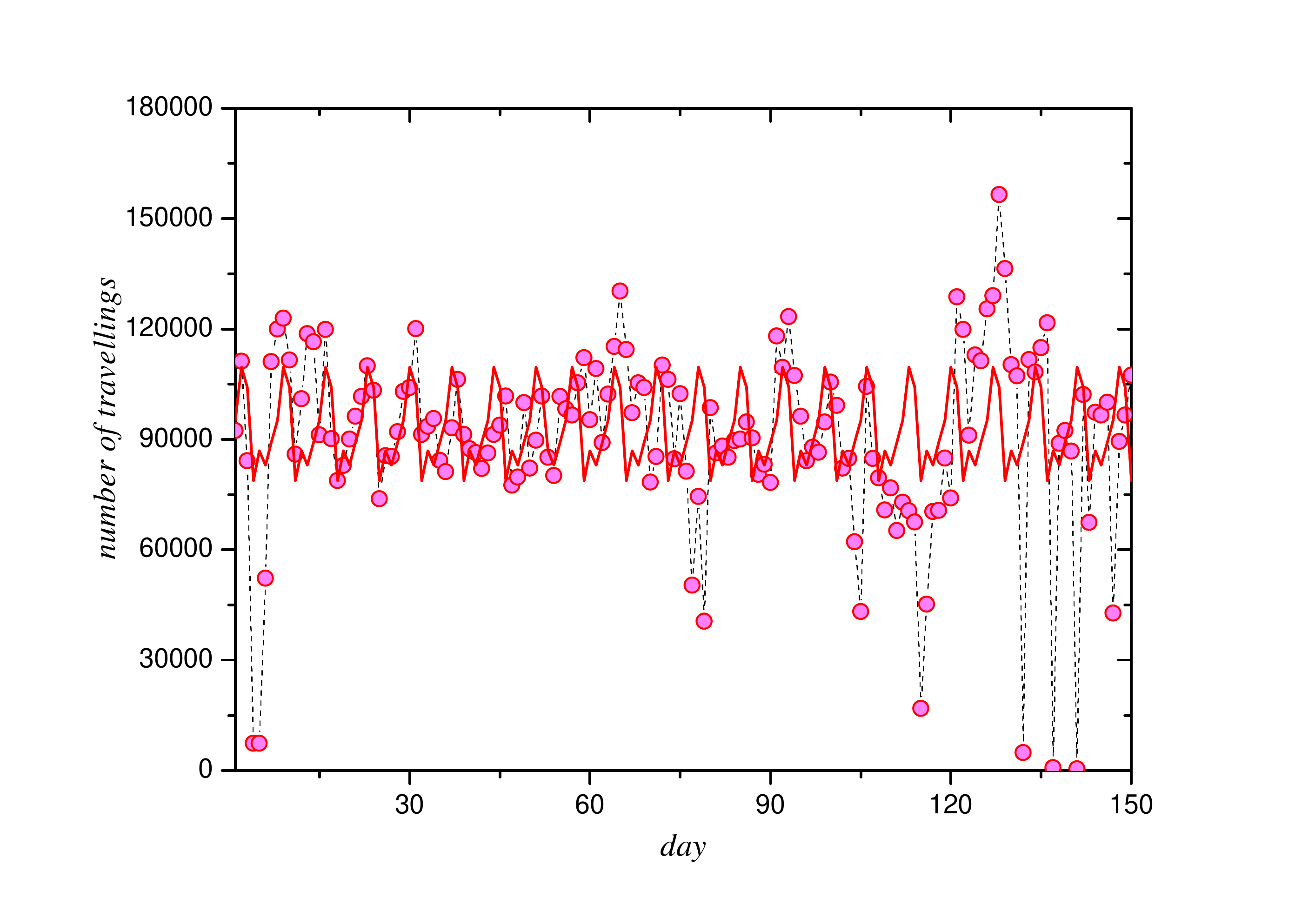}
\caption{The volume of mobility between sub-prefectures for each day. The circles show the empirical human mobility volume, the dashed line are inserted for eye guidance. The red line is the periodical tendency obtained by averaging the data on each day of the week. The relative low value of the mobility volume might be due to terrible weather conditions, serious social disturbances or technical reasons related to collecting data \cite{D4D2012Cote}.}\label{fig.volume}
\end{figure}

With the Call Detailed Records (CDR) data from 12/01/2011 to 04/28/2012, a random sample of 0.5 million cellphone users within C\^{o}te d'Ivoire was provided by the Orange Group and the organizers of the D4D Challenge \cite{D4D2012Cote,website:OrangeInc}. The CDR contains anonymized user-id, timestamp and subpref-id (the coordinates of which are also provided) \cite{D4D2012Cote}, and therefore, we can extract the mobility traces of any user by focusing on his (or her) location changes for any day. In this paper, we set the time window to be one day: within one day, if one user appears at location $M$ and then $N$, we deduce that a movement from $M$ to $N$ occurred (an inevitable drawback for extracting mobility from CDR is that only when the user has an activity can his/her position be recorded. Therefore, there will certainly be some missing mobility traces). We also filter the pass-by nodes according to the observed stay duration \cite{jiang2013review} in a sub-prefecture (i.e., if someone appear in $M$ then $N$ and then $P$, but he stays in $N$ for quite a short time, then $N$ is highly not be his destination, we will identify $M$ to $P$ as a movement rather than $M$ to $N$ and $N$ to $P$). Because we mainly focus on national epidemic spreading, we extracted the mobility matrix $A$ at the sub-prefecture resolution (i.e., treated a sub-prefecture as a node in the network). The volume of the human mobility between sub-prefectures for each day is shown in Fig. \ref{fig.volume}.

\begin{figure}   \centering
\includegraphics[width=3.5in]{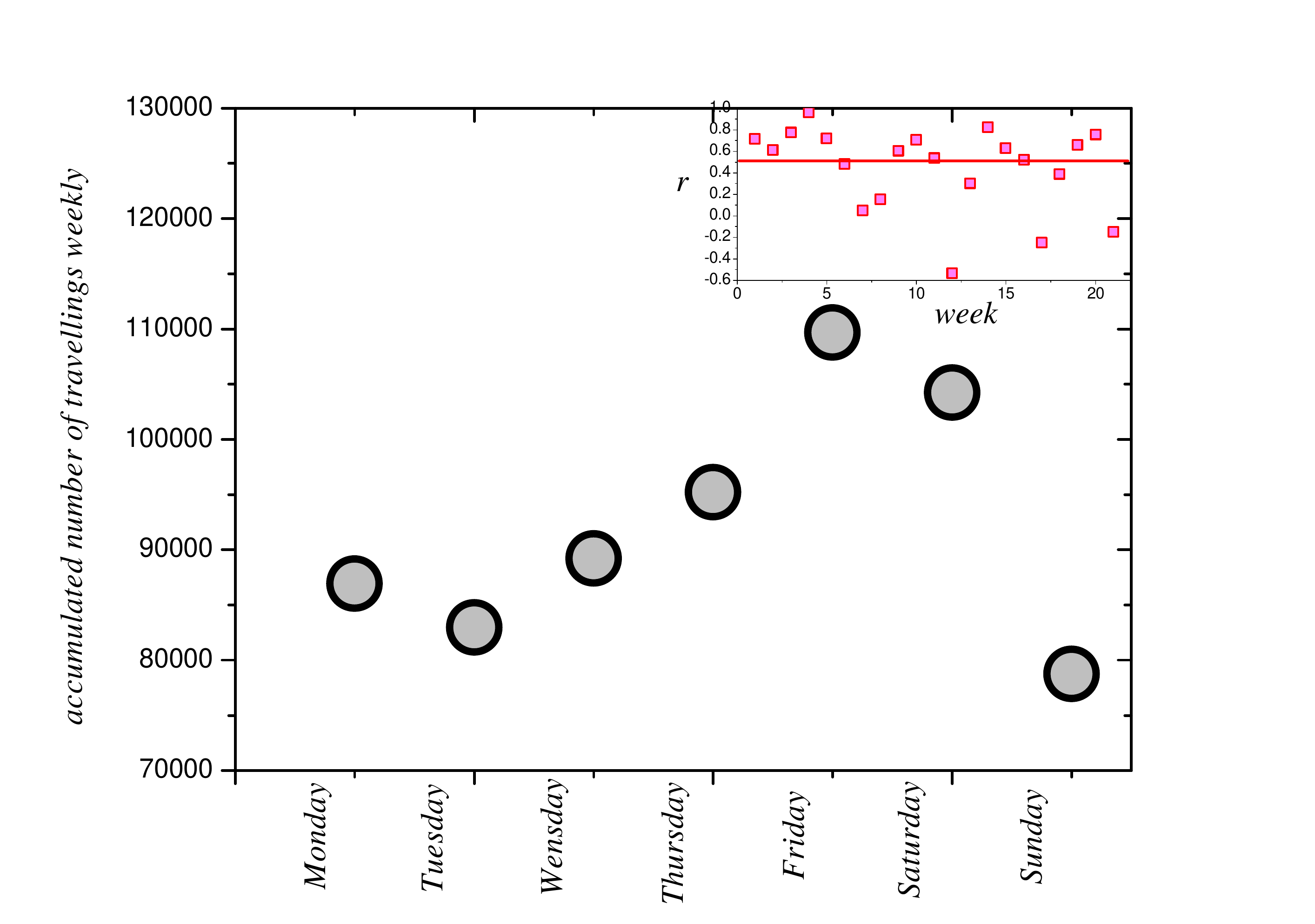}
\caption{The average volume of the mobility inter-sub-prefecture for each day in a week for 150 days. (inset) The Pearson correlation between the averaged data and real data.}\label{fig.volumeWeek}
\end{figure}

When we accumulate the mobility volume by day of the week, we observe a  periodical property (see Fig. \ref{fig.volumeWeek}), which might be mainly affected by the commuting activities. Friday shows the peak volume, while the weekend shows comparatively low volumes. The Pearson correlations between this periodical change and the real data are very strong (most of them are larger than 0.5; see the inset of Fig. \ref{fig.volumeWeek}) apart from some noise and missing data due to technical (about 100 hours of data are missing) and social issues \cite{D4D2012Cote}. In the simulation, the time step corresponds to one day, and we use the real time mobility volume data for each day.

\section{Supplementary Figures}
\begin{figure} [htbp]  \centering
\includegraphics[width=4.5in]{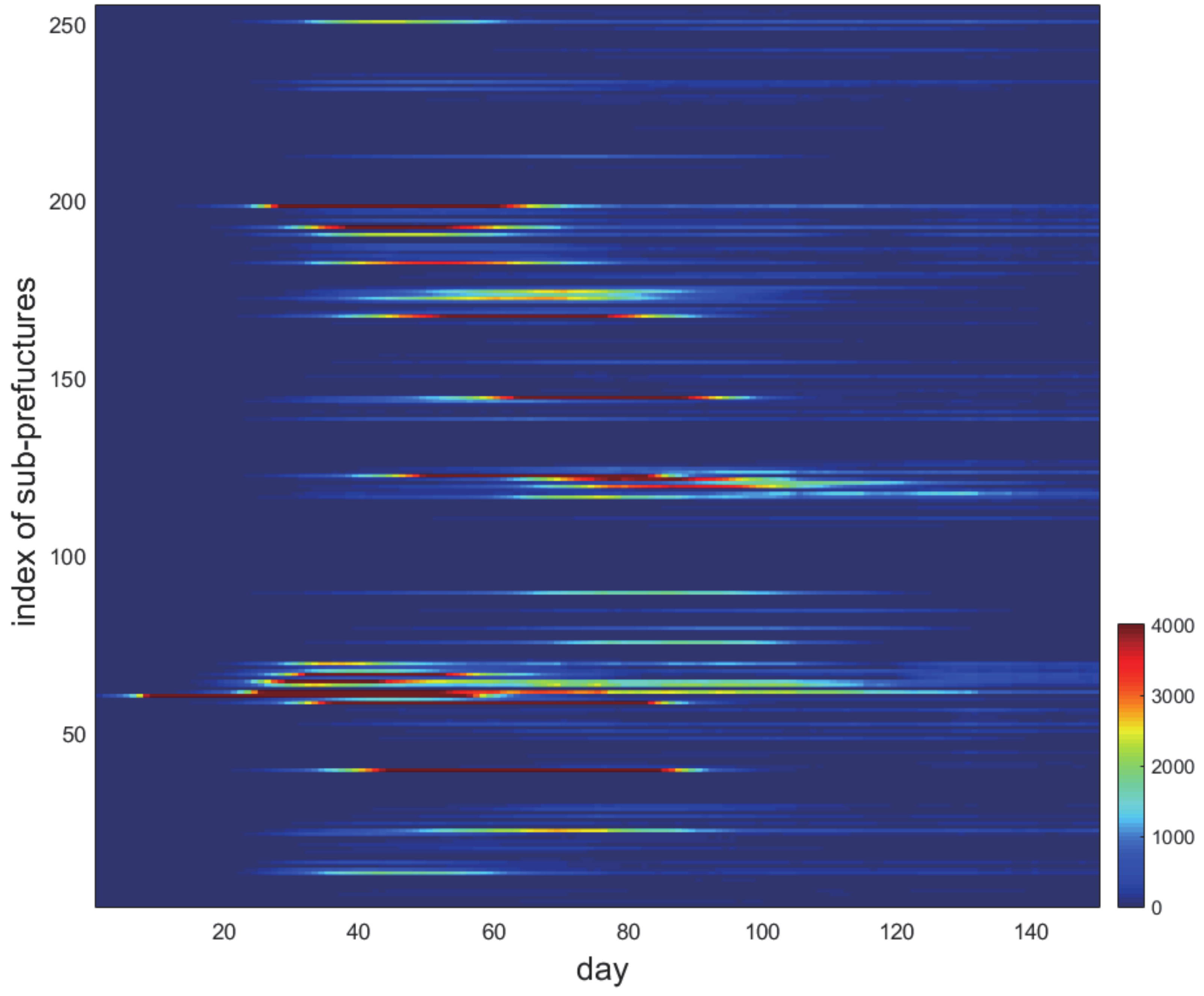}
\caption{The unsorted temporal epidemic spreading process. The $X$ axis is the time (i.e., day), and the $Y$ axis is the index of the sub-prefectures.}\label{fig.Apcolor}
\end{figure}

\begin{figure} [htbp]  \centering
\includegraphics[width=5in]{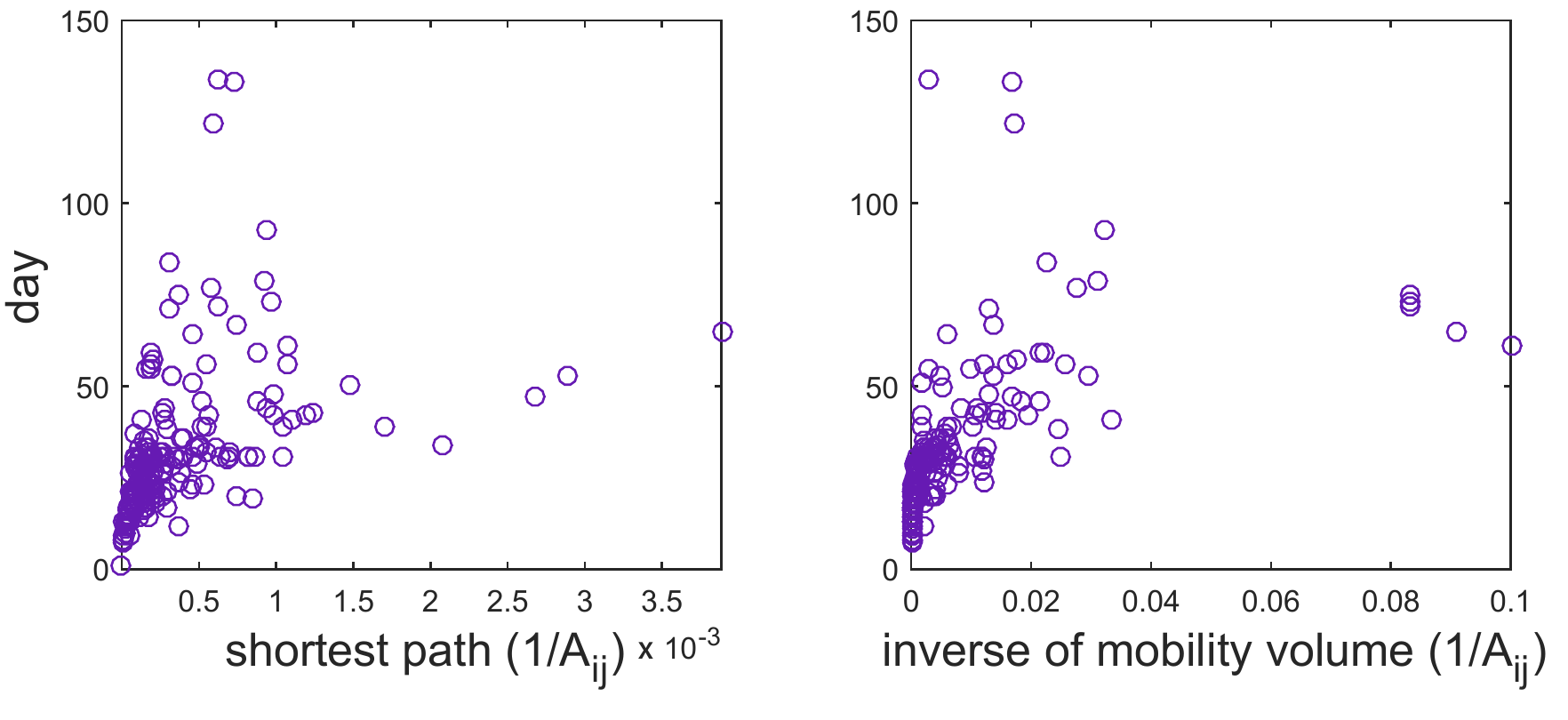}
\caption{The full picture of Fig. \ref{fig.scatter}(b) and (c).}\label{fig.Ascatter}
\end{figure}


\end{document}